\documentclass[conference,a4paper]{IEEEtran}

\usepackage[utf8]{inputenc} 
\usepackage[T1]{fontenc}
\usepackage{url}
\usepackage{ifthen}
\usepackage{cite}
\usepackage[cmex10]{amsmath} 
\usepackage{amssymb}
\usepackage{graphicx}
\usepackage{multirow}
\usepackage{array}
\usepackage{makecell}
\usepackage{tikz}
\usepackage{algorithm}
\usepackage{algpseudocode}
\usepackage{comment}
\usepackage[english]{babel}

\newtheorem{thm}{Theorem}
\newtheorem{lem}{Lemma}

\newtheorem{defn}{Definition}

\newtheorem{exmp}{Example}

\begin{document}
\title{Multi-access Coded Caching Schemes From Cross Resolvable Designs} 
	
\author{%
		\IEEEauthorblockN{Digvijay Katyal, Pooja Nayak M and B. Sundar Rajan \IEEEauthorrefmark{1}}
		\IEEEauthorblockA{\IEEEauthorrefmark{1}Department of Electrical Communication Engineering, Indian Institute of Science, Bengaluru 560012, KA, India \\
			E-mail: \{digvijayk,poojam,bsrajan\}@iisc.ac.in}
}
\maketitle
\begin{abstract}
We present a novel caching and coded delivery scheme for a multi-access network where multiple users can have access to the same cache (shared cache) and any cache can assist multiple users. This scheme is obtained from resolvable designs satisfying certain conditions which we call {\it cross resolvable designs}. To be able to compare different multi-access coded schemes with different number of users we normalize the rate of the schemes by the number of users served. Based on this per-user-rate we show that our scheme performs better than the well known Maddah-Ali - Nieson (MaN) scheme and the recently proposed ("Multi-access coded caching : gains beyond cache-redundancy" by Serbetci, Parrinello and Elia) SPE scheme. It is shown that the resolvable designs from affine planes are cross resolvable designs and our scheme based on these performs better than the MaN scheme for large memory size cases. The exact size beyond which our performance is better is also presented. The SPE scheme considers only the cases where the product of the number of users and the normalized cache size is 2, whereas the proposed scheme allows different choices depending on the choice of the cross resolvable design. 
\end{abstract}
\section{INTRODUCTION}
\label{sec1}	
Caching techniques help to reduce data transmissions during the times of high network congestion by prefetching parts of popularly demanded contents into the memories of end users. The seminal work of \cite{MaN} provided a coded delivery scheme which performed within a constant factor of the information-theoretic optimum for all values of the problem parameters.

The idea of a placement delivery array $($PDA$)$ to represent the placement and delivery phase of a coded caching problem first appeared in \cite{YCTC}.  These PDAs could represent any coded caching problem with symmetric prefetching and the popular Ali-Niesen scheme was also found to be a special case. Since then many different constructions of PDA have been put forth achieving low subpacketization levels\cite{TaR}.

Recently, placement delivery arrays have found applications in different variants of coded caching scenarios like in Device to device (D2D) networks as D2D placement delivery array (DPDA) and in Combination networks as combinational PDA (C-PDA) \cite{WCYT,YWY}.	

Most of the works on coded caching consider scenarios where each user has its own dedicated cache. However in a variety of settings, such as different cellular networks, multiple users share a single cache or users  can conceivably connect to multiple caches whose coverage areas may overlap. The possibility of users to access more than one cache was first addressed in \cite{HKD}. This was motivated by the upcoming heterogeneous cellular architecture which will contain a dense deployment of wireless access points with small coverage and relatively large data rates, in addition to sparse cellular base stations with large coverage area and small data rates. Placing cache at local access points could significantly reduce the base station transmission rate, with each user being able to access content at multiple access points along with the base station broadcast. The work in \cite{HKD}  considered a K-user shared-link broadcast channel where each user is assisted by exactly $L  > 1$ caches (with a cyclic wrap around), and where each cache can serve exactly $L$ users. The authors called this as multi-access problem and derive an achievable rate and information theoretic lower bound which differs by a multiplicative gap scaling linearly with $L$. Later in \cite{RaK,RaK2}, new bounds for the optimal rate-memory trade-off were derived for the same problem. Also a new achievable rate for general multi-access setup which is order-wise better than the rate in \cite{HKD} is derived. The authors focus on the special case with $L\geq K/2$ and provide a general lower bound on the optimal rate and establish its order optimal memory rate trade off under the restrictions of uncoded placement. For few special cases like $L=K-1;\;L=K-2;\;L=K-3$ with $K$ even, exact optimal uncoded  memory-rate trade off is derived.
	
The work of Serbetci et al.\cite{SPE} gives yet another class of coded caching schemes for the multi-access setup, where each user in a $K$-user shared link broadcast channel is connected to $z>1$ caches (with a cyclic wrap around), and where each cache can serve exactly $z$ users. This was the first instance where authors have analyzed this scenario in the context of worst case delivery time and when the number of files in server database is greater than equal to the number of users, the proposed scheme experiences a larger gain than in \cite{MaN}.
	
In \cite{SBP2} authors consider the shared cache scenario where multiple users share the same cache memory, and each user is connected to only one cache. The work in \cite{SBP2} considers the shared link network, with $K$ users and $\Lambda,\;\Lambda\leq K$ helper caches, where each cache is assisting arbitrary number of distinct users. Also each user is assigned to single cache. For this set up, the authors identify the fundamental limits under the assumption of uncoded placement and any possible user to cache association profile. The authors derive the exact optimal worst-case delivery time.

In \cite{AAA}, the authors propose a coded placement scheme for the setup where the users share the end-caches and showed that the proposed scheme outperforms the scheme in \cite{SBP2}. In this scheme the authors use both coded and uncoded data at the caches, taking into consideration the users connectivity pattern. Firstly, for a two-cache system, authors provided an explicit characterization of the gain from coded placement, then the scheme is extended to $L-$cache systems, where authors obtain optimal parameters for the caching scheme by solving a linear program.
	
The schemes so far mentioned above in context of shared caches/multi-access scenarios consider the special framework with cyclic wrap around, ensuring that intersection of all caches that a user is connected to is empty. In \cite{BL}, the authors addressed a system model involving a cache sharing strategy where the set of users served by any two caches is no longer empty. In \cite{BL}, the authors consider the caching problem with shared caches, consisting of a server connected to users through a shared link, where a pair of users share two caches. 
\begin{figure}
\begin{center}
\includegraphics[width=7cm,height=6cm]{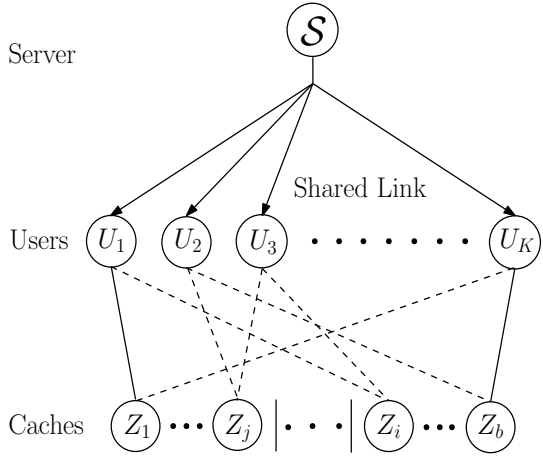}
\caption {Problem setup for multi-access coded caching with $K$ users, $b$ caches and each user, connected to $z$ caches.}
\label{fig1}
\end{center}
\end{figure}

Various combinatorial designs have been used in different setups of coded caching \cite{TaR2,KrP1,KrP2}. To the best of our knowledge this is the first work that uses designs for multi-access coding caching.
\subsection{Multi-access Coded Caching - System Model}
\label{sec1A}
Fig. \ref{fig1} shows a multi-access coded caching system with a unique server $\mathcal{S}$ storing $N$ files $W_{1}$,$W_{2}$,$W_{3}$,\dots,$W_{N}$ each of unit size. There are $K$ users in the network  connected via an error free shared link to the server $\mathcal{S}.$ The set of users is denoted by $\mathcal{K}.$  There are $b$ number of helper caches each of size $M$ files. Each user has access to $z$ out of the $b$ helper caches. Let $\mathcal{Z}_k$ denotes the content in the $k$-th cache. It is  assumed that each user has an unlimited capacity link to the caches it is connected to. 

There are two phases: the placement phase and the delivery phase. During the placement phase certain parts of each file are stored in each cache which is carried out during the off-peak hours. During the peak hours each user demands a file and the server broadcasts  coded transmissions such that each user can recover its demand by combining the received transmissions with what has been stored in the caches it has access to. This is the delivery phase. The coded caching problem is to jointly design the placements and the delivery with minimal number of transmissions to satisfy the demand of all the users. The amount of transmissions used in the unit of files is called the {\it rate} or the {\it delivery time}. Subpacketization level is the number of packets that a file is divided into. Coding gain is defined as the number of users benefited in a transmission. 
\subsection{The Maddah-Ali Nieson (MaN) Coded Caching Scheme}
\label{sec1B}
The framework of the seminal paper \cite{MaN} considers a network with $K$ users, each equipped with memory of size ${M}$ and $N$ files of very large size among which each user is likely to demand any one file. The rate $R$ achieved is  
       $$ K  \left(1 - \frac{{M}}{N}\right) \frac{1}{(1 + K\frac{{M}}{N})}.$$
The factor $(1 + K\frac{{M}}{N})$ which is originally call global caching gain is also known as the {\it coding gain} or the {\it Degrees of Freedom (DoF)}. We refer to this scheme as the MaN scheme henceforth. This original setup can be viewed as a special case of the scheme corresponding to Fig. \ref{fig1} with $b=K$ and $z=1$ which may be viewed as each user having a dedicated cache of its own. 
\subsection{Serbetci-Parrinello-Elia (SPE) Multi-access Coded Caching Scheme}
\label{sec1C}
In \cite{SPE}, a network consisting of $K$ users connected via an error free shared link to a server $\mathcal{S}$ storing $N$ files $W_{1},W_{2},W_{3},\dots,W_{N}$ is considered. Each user in the network can access $z$ caches out of $K$ helper caches, each of size $M$ = $N\gamma$ $($units of file$)$, where $\gamma \in \{\frac{1}{K},\frac{2}{K},\dots, 1\}$. The setup of this scheme, which we refer henceforth as the SPE scheme, can be considered as a special case of the setup shown in Fig.\ref{fig1} where  each user $k$ $\in$ $\mathcal{K}$ is associated with the caches,
	$$C_{k} := {(k, k+1 ,\dots,k+z-1)_{z}} \subseteq \mathcal{K.}$$
In \cite{SPE} the authors focus on the special case, $K\gamma = 2$ and provide a scheme  which exceed the Ali-Niesen coding gain $K\gamma + 1$.  Also, for the special case with access to an integer number $z$ = $ \frac{K - 1}{K\gamma}$ of caches of normalized size $\gamma$, the optimal rate taking the form $ K \frac{1-\gamma z}{K\gamma z+ 1} = \frac{1}{K}$ 
corresponding to a degrees of freedom (DoF) of $K\gamma z + 1$ users served at a time is reported.
\subsection{Comparing different multi-access coded caching schemes}
\label{sec1D}
In any Multi-access Coded caching problem the design parameters are the number of files $N$, number of users $K$, number of caches $b$, the memory size of the cache, $M$ files and number of caches a user has access to, $z$. For two Multi-access schemes the number of users may be different depending upon the cache-user topology, keeping other parameters the same. So comparing such two schemes with respect to rate $R$  the existing rate $R$ may be misleading. Therefore to compare our multi-access scheme with other existing schemes we introduce a new term that normalizes the existing rate $R$ with the number of users $K$ supported i.e rate per user or per user rate $\frac{R}{K}$. The lower the rate per user for a given $b, M, N, z$ the better the scheme.


\subsection{Contributions}
\label{sec1E}
The contributions of this paper may be summarized as follows:
\begin{itemize}
\item A subclass of resolvable designs (called cross resolvable designs) is identified using which new classes of multi-access coding schemes are presented. 
\item  To be able to compare different multi-access coded schemes with different number of users we normalize the rate of the schemes by the number of users served. Based on this per-user-rate we show that our scheme performs better than the MaN scheme and the SPE scheme for several cross resolvable designs. 
\item It is shown that the resolvable designs from affine planes \cite{Stinson} are cross resolvable designs and our scheme performs better than the MaN scheme for large memory size cases. The exact size beyond which our performance is better is also presented. 
\item The resolvable designs from affine planes are a special case of the affine resolvable designs from balanced incomplete block designs (AR-BIBDs) parametrized by an integer $m \geq 2,$ which are CRDs. The case affine resolvable designs from affine planes corresponds to $m=2.$ It is shown that the advantage in terms of subpacketization holds for all values of $m$ the advantage in terms of per user rate holds only for $m=2$ and $m=3$ for large memory sizes.  

\item The SPE scheme \cite{SPE} considers only the cases where $\frac{KM}{N} = 2$, while the proposed scheme allows different choices of $\frac{KM}{N}$ depending on the choice of the cross resolvable design.
\end{itemize}
        The paper is organized as follows. Section \ref{sec2} describes all the details related to  resolvable designs and defines a subclass of resolvable designs termed in this paper as cross resolvable designs (CRDs). Our proposed scheme associated with CRDs is described in Section \ref{sec3}. Comparison of performance of our scheme with the MaN and the SPE schemes constitute Section \ref{sec4}.  In Section \ref{sec5} we discuss schemes obtained from resolvable designs for which the number of caches is equal to the number of users. This includes coded caching schemes from resolvable designs that are not cross resolvable.  Concluding remarks constitute Section \ref{sec6} and the proof of correctness of our delivery algorithm is given in the Appendix.
\section{CROSS RESOLVABLE DESIGNS}
\label{sec2}
We use a class of combinatorial designs called resolvable designs\cite{Stinson} to specify placement in the caches.
\begin{defn}\cite{TaR}
A design is a pair  $(X, \mathcal{A})$ such that
\begin{itemize}
\item $X$ is a finite set of elements called points, and
\item $\mathcal{A}$ is a collection of nonempty subsets of $X$ called blocks, where each block contains the same number of points.
\end{itemize}
\end{defn}
\begin{defn}\cite{TaR}
A parallel class $\mathcal{P}$ in a design $(X, \mathcal{A})$ is a subset of disjoint blocks from $\mathcal{A}$ whose union is $X$. A partition of $\mathcal{A}$ into several parallel classes is called a resolution, and $(X, \mathcal{A})$ is said to be a resolvable design if A has at least one resolution.
\end{defn}
\begin{exmp}
\label{exmp1}
Consider a block design specified as follows.
\begin{align*}
X =& \{1, 2, 3, 4\}, \text{ and}\\
\mathcal{A} = &\{\{1, 2\},\{1, 3\},\{1, 4\},\{2, 3\},\{2, 4\},\{3, 4\}\}.
\end{align*}
It can be observed that this design is resolvable with the following parallel classes.
\begin{align*}
\mathcal{P}_1 =& \{\{1, 2\},\{3, 4\}\},\\
\mathcal{P}_2 =& \{\{1, 3\},\{2, 4\}\}, \text{ and}\\
\mathcal{P}_3 =& \{\{1, 4\},\{2, 3\}\}.
\end{align*}

Note that in above example, $\mathcal{P}_1$, $\mathcal{P}_2$, $\mathcal{P}_3$ forms a partition of $\mathcal{A}$. If $\mathcal{A}$ = \{\{1, 2\},\{1, 3\},\{3, 4\},\{2, 4\}\}, we get another resolvable design with two parallel classes $\mathcal{P}_1$ and $\mathcal{P}_2$.
\end{exmp}
\begin{exmp}
\label{exmp2}
Consider a block design specified as follows.
\begin{align*}
X =& \{1, 2, 3, 4, 5, 6\}, \text{ and}\\
\mathcal{A} = &\{\{1, 2 ,3\},\{4, 5, 6\},\{1, 4 ,5\},\{2, 3, 6\}\}.
\end{align*}
It can be observed that this design is resolvable with the following parallel classes.
\begin{align*}
\mathcal{P}_1 =&\{\{1, 2 ,3\},\{4, 5, 6\}\}\\
\mathcal{P}_1 =&\{\{1, 4 ,5\},\{2, 3, 6\}\}
\end{align*}
\end{exmp}
For a given resolvable design  $(X, \mathcal{A})$ if |$X$| = $v$, |$\mathcal{A}$| = $b$ , block size is $k$ and number of parallel classes is $r$, then there are exactly $\frac{b}{r}$ blocks in each parallel class. Since the blocks in each parallel class are disjoint, therefore number of blocks in each parallel class = $\frac{b}{r}$ = $\frac{v}{k}$.
\subsection{Cross Resolvable Design (CRD)}
	
\begin{defn}[\textbf{Cross Intersection Number}]
For any resolvable design $(X, \mathcal{A})$ with $r$ parallel classes, the $i^{th}$ cross intersection number, $\mu_{i}$ where $i \in \{2, 3, \dots,r\}$, is defined as the cardinality of intersection of $i$ blocks drawn from any $i$ distinct parallel classes, provided that, this value remains same ($\mu_i\neq 0$), for all possible choices of blocks. \\
		For instance, in Example 1, $\mu_{2}$ = 1, as the intersection of any 2 blocks drawn from 2 distinct parallel classes is always at exactly one point. But we cannot define $\mu_{3}$ as the intersection of $3$ blocks drawn from 3 distinct parallel classes takes elements from the set \{0, 1\}.
\end{defn}
\begin{defn}[\textbf{Cross Resolvable Design}]
For any resolvable design $(X, \mathcal{A})$, if there exist at least one $i\in \{2,3,\dots,r\}$ such that the $i^{th}$ cross intersection number $\mu_{i}$ exists, then the resolvable design is said to be a Cross Resolvable Design (CRD). For a CRD the maximum value for $i$ for which $\mu_i$ exists is called the Cross Resolution Number (CRN) for that CRD. A CRD  with the CRN equal to $r$ is called a Maximal Cross Resolvable Design (MCRD).
\end{defn}

Note that the resolvable design in \textit{Example 2} is not a CRD  as $\mu_{2}$ does not exist.
\begin{exmp}
\label{exmp3}
For the resolvable design $(X, \mathcal{A})$ with
\begin{align*}
X = &\;\{1, 2, 3, 4, 5, 6, 7, 8, 9\}, \text{ and} \\
\mathcal{A} =& \;\{\{1, 2, 3\},\{4, 5, 6\},\{7, 8, 9\},\\
&\{1, 4, 7\},\{2, 5, 8\},\{3, 6, 9\}\},
\end{align*}
the parallel classes are
\begin{align*}
\mathcal{P}_1 = &\;\{\{1, 2, 3\},\{4, 5, 6\},\{7, 8, 9\}\}, \text{ and} \\
\mathcal{P}_2 =&\; \{\{1, 4, 7\},\{2, 5, 8\},\{3, 6, 9\}\}.
\end{align*}
It is easy to verify that  $\mu_{2} = 1$.	
\end{exmp}
\begin{exmp}
\label{exmp4}
For the resolvable design $(X, \mathcal{A})$ with 
\begin{align*}
X =&\; \{1, 2, 3, 4, 5, 6, 7, 8\}, \text{ and} \\
\mathcal{A} =&\; \{\{1, 2, 3, 4\},\{ 5, 6, 7, 8\},\{1, 2, 5, 6\},\\
&\{3, 4, 7, 8\},\{1, 3, 5, 7\},\{2, 4, 6, 8\}\},
\end{align*}
the parallel classes are
\begin{align*}
\mathcal{P}_1 =&\; \{\{1, 2, 3, 4\},\{5, 6, 7, 8\}\}, \\
\mathcal{P}_2 =&\; \{\{1, 2, 5, 6\},\{3, 4, 7, 8\}\}, \text{ and} \\
\mathcal{P}_3 =&\; \{\{1, 3, 5, 7\},\{2, 4, 6, 8\}\}.
\end{align*}
In this case  $\mu_{2}$ = 2 and $\mu_{3}$ = 1.
\end{exmp}	
\begin{exmp}
\label{exmp5}
Consider the resolvable design $(X, \mathcal{A})$ with
\begin{align*}
X = &\;\{1, 2, 3, 4, 5, 6, 7, 8, 9, 10, 11, 12\}, \text{ and} \\
\mathcal{A} =&\; \{\{1, 2, 3, 4, 5, 6\},\{ 7, 8, 9, 10, 11, 12\},\\
&\{1, 2, 3, 7, 8, 9\},\{ 4, 5, 6, 10, 11, 12\}\}.
\end{align*}
The parallel classes are
\begin{align*}
\mathcal{P}_1 = &\;\{\{1, 2, 3, 4, 5, 6\},\{ 7, 8, 9, 10, 11, 12\}\}, \\
\mathcal{P}_2 =&\; \{\{1, 2, 3, 7, 8, 9\},\{ 4, 5, 6, 10, 11, 12\}\}.
\end{align*}
We have $\mu_{2}$ = 3.
\end{exmp}
\begin{exmp}
    \label{exmp6}
    Consider the resolvable design $(X, \mathcal{A})$ with
    \begin{align*}
    X = &\;\{1, 2, 3, 4, 5, 6, 7, 8, 9\}, \text{ and} \\
    \mathcal{A} =& \;\{\{1, 2, 3\},\{4, 5, 6\},\{7, 8, 9\},\{1, 4, 7\},\{2, 5, 8\},\{3, 6, 9\},\\
    &\;\{1, 5, 9\},\{2, 6, 7\},\{3, 4, 8\},\{1, 6, 8\},\{2, 4, 9\},\{3, 5, 7\}\}.
    \end{align*}
    The parallel classes are
    \begin{align*}
    \mathcal{P}_1 = &\;\{\{1, 2, 3\},\{4, 5, 6\},\{7, 8, 9\}\},\\
    \mathcal{P}_2 =&\; \{\{1, 4, 7\},\{2, 5, 8\},\{3, 6, 9\}\},\\
    \mathcal{P}_3 =&\; \{\{1, 5, 9\},\{2, 6, 7\},\{3, 4, 8\}\},\text{ and}\\
    \mathcal{P}_4 =&\; \{\{1, 6, 8\},\{2, 4, 9\},\{3, 5, 7\}\}.
    \end{align*}
    Here  $\mu_{2} = 1$ and $\mu_{3}$, $\mu_{4}$ does not exist. 
\end{exmp}

From Example \ref{exmp6} one can see that $\mu_{r}$ need not always exist for a CRD.
\begin{lem}
\label{lem1}
For any given CRD $(X, \mathcal{A})$ with $r$ parallel classes and any cross intersection number $\mu_{i}$ for $i \in \{3,4,\dots,r\}$ we have,
$$\mu_{i-1} = \mu_{i}\frac{v}{k}$$
\end{lem}
\begin{IEEEproof}
From any $i$ parallel classes, let us choose a block from each parallel class denoted as $\mathcal{A}_{1,1},\mathcal{A}_{2,1},\mathcal{A}_{3,1},\dots,\mathcal{A}_{i,1}$.
Let $$\mathcal{A}_{1,1} \cap \mathcal{A}_{2,1} \cap \mathcal{A}_{3,1} \cap \dots \cap \mathcal{A}_{i,1} = \mathcal{M}_1$$
$$\mathcal{A}_{1,1} \cap \mathcal{A}_{2,1} \cap \mathcal{A}_{3,1} \cap \dots \cap \mathcal{A}_{i,2} = \mathcal{M}_2$$
$$\vdots$$
$$\mathcal{A}_{1,1} \cap \mathcal{A}_{2,1} \cap \mathcal{A}_{3,1} \cap \dots \cap \mathcal{A}_{i,\frac{v}{k}} = \mathcal{M}_\frac{v}{k}$$
		
From definition of cross resolvable design,
$$|\mathcal{M}_1| =| \mathcal{M}_2|=| \mathcal{M}_3|= \dots =|\mathcal{M}_\frac{v}{k}|=\mu_{i}$$

It is also easy to see that, any $\mathcal{M}_k \cap \mathcal{M}_l = \phi$ where $k,l \in \{1,2,\dots,\frac{v}{k}\}$ 
Now,
\begin{multline*}
		\mathcal{M}_1 \cup \mathcal{M}_2  \cup \dots \cup \mathcal{M}_\frac{v}{k} = \\\mathcal{A}_{1,1} \cap \mathcal{A}_{2,1} \cap \mathcal{A}_{3,1} \cap \dots \cap \mathcal{A}_{i-1,1}
		\cap\left\{\underset{l\in[\frac{v}{k}]}\bigcup{\mathcal{A}_{i,l}}\right\}
\end{multline*}
$$=\mathcal{A}_{1,1} \cap \mathcal{A}_{2,1} \cap \mathcal{A}_{3,1} \cap \dots \cap \mathcal{A}_{i-1,1}$$
since $\underset{l\in[\frac{v}{k}]}\bigcup\mathcal\{{A}_{i,l}\} = \mathcal{X}$. We have
$$|\mathcal{M}_1 \cup \mathcal{M}_2  \cup \dots \cup \mathcal{M}_\frac{v}{k}| = \mu_{i}\frac{v}{k}$$
$$|\mathcal{A}_{1,1} \cap \mathcal{A}_{2,1} \cap \mathcal{A}_{3,1} \cap \dots \cap \mathcal{A}_{i-1,1}| = \mu_{i}\frac{v}{k}$$
$$\mu_{i-1} = \mu_{i}\frac{v}{k}.$$
\end{IEEEproof}
\section{PROPOSED SCHEME}
\label{sec3}
Given a cross resolvable design $(X, \mathcal{A})$ with $v$ points, $r$ parallel classes, $b$ blocks of size $k$ each, $b_r\stackrel{def}{=}\frac{b}{r}$ blocks in each parallel class, we choose some $z \in \{2,3,\dots,r\}$ such that $\mu_{z}$ exists. Let $\mathcal{A}_j$ denote the $j^{th}$ block in $\mathcal{A}$, assuming some ordering on the blocks of $\mathcal{A}$. We associate  a coded caching problem with $K = {r \choose z} (\frac{b}{r})^z$ number of users, $N$ files in server database, $b$  number of caches, $\frac{M}{N} = \frac{k}{v}$ fraction of each file at each cache and subpacketization level $v.$  A user is connected to distinct $z$ caches such that these $z$ caches correspond to $z$ blocks from distinct parallel classes. 
 We denote the set of $K$ users  $\mathcal{K}$ as,
$$\mathcal{K} := \{U_{H} :\;|H| = z\}$$
where, $H$ is a $z$ sized set containing cache indices from distinct parallel classes.
\subsection{Placement Phase}
\label{sec3A} 
In the caching placement phase, we split each file $W_i,\;\forall i \in [N]$ into $v$ non-overlapping subfiles of equal size i.e.
	$$W_i = (W_{i,k} : \forall k \in [v]),\; i = 1,2,\dots,N.$$
	
The placement is as follows. In the $j^{th}$ cache, the indices of the subfiles stored in $\mathcal{Z}_j$ is the $j^{th}$ block in the design. We assume symmetric batch prefetching i.e.,
$$\mathcal{Z}_j = \{{W}_{ik} : k \in \mathcal{A}_j\},\; \forall i \in \{1,2...,N\}, \forall j \in \{1,2...,b\}$$
Therefore the total number of subfiles for each file in any cache is block size $k$ of the resolvable design i.e. $\frac{M}{N}=\frac{k}{v}$.  

Let $M'$ denote the size of the memory in units of files that a user has access to. We have
\begin{multline*}
	\frac{M'}{N}= \sum_{i=1}^z \frac{|\mathcal{A}_i|}{v}-\sum_{1\leq i_1<i_2\leq z}^z \frac{|\mathcal{A}_{i_1}\cap\mathcal{A}_{i_2}|}{v}+\dots +(-1)^{t+1}\\\sum_{1\leq i_1<\dots<i_t\leq z}^{z} \frac{|\mathcal{A}_{i_1}\cap\dots \cap\mathcal{A}_{i_t}|}{v}
	+\dots+ (-1)^{z+1}\frac{|\mathcal{A}_{1}\cap\dots\cap\mathcal{A}_{z}|}{v}
	\end{multline*}
	where $\mathcal{A}_i,\;i\in[z]$ are $z$ blocks from $z$ distinct parallel classes. Using Lemma \ref{lem1} we get,   \\
	\begin{multline*}
	\frac{M'}{N}= \sum_{i=1}^z \frac{M}{N}-\sum_{1\leq i_1<i_2\leq z}^z \frac{\mu_2}{v}+\dots +(-1)^{t+1}\sum_{1\leq i_1<\dots<i_t\leq z}^{z} \frac{\mu_t}{v}\\
	+\dots+ (-1)^{z+1}\frac{\mu_z}{v},
	\end{multline*}
which simplifies to
	$$\frac{M'}{N}= \frac{zM}{N}+\sum_{t=2}^z (-1)^{t+1}\binom{z}{t}\frac{\mu_t}{v}.$$
From the above expression it is clear that for cross resolvable design $M'\neq zM$. All the cases considered in \cite{SPE,RaK3} corresponds to the case that $M'= zM,$ i.e., the size of the memory that a user has access to is an integer multiple of the size of the cache. From this it follows that the cases considered in \cite{SPE} do not intersect with the cases considered in this paper. 
\begin{lem}
\label{lem2}
		The number of users having access to any particular subfile is exactly
		$\binom{r}{z}(b_r^z-(b_r-1)^z)$.
\end{lem}
\begin{IEEEproof}
		There are $\binom{r}{z}$ possible ways of choosing $z$ parallel classes out of $r$ parallel classes. Fix some point $l \in [v]$. In each parallel class there will be $(b_r-1)$ blocks that does not contain $l$. So there are totally $(b_r-1)^z$ users which do not have access to $l$. Hence the number of users which have access to $l$ is given by $(b_r^z-(b_r-1)^z)$. So, taking all possible combinations of $z$
		parallel classes we have
		$\binom{r}{z}(b_r^z-(b_r-1)^z).$
\end{IEEEproof}
\begin{lem}
\label{lem3}
		The number of users having access to a subfile of a specific cache is exactly $\binom{r-1}{z-1}(b_r)^{z-1}$.
\end{lem}
\begin{IEEEproof}
		Any user is connected to $z$ parallel classes. First we fix a subfile accessible to the user by fixing a cache accessible to the user. Once we fix a cache we also fix a parallel class. Then other $z-1$ parallel classes can be chosen in $\binom{r-1}{z-1}$ ways. Since there are $b_r$ blocks in each of these parallel classes, we have total number of users equal to 
		$\binom{r-1}{z-1}(b_r)^{z-1}.$
	\end{IEEEproof}
\subsection{Delivery Phase}
\label{sec3B}
For delivery, we first arrange the users in lexicographical order of their indices $S$, establishing a one-to-one correspondence with the set $\{1, 2, \dots , K \}$.
	At the beginning of the delivery phase, each user requests one of the $N$ files and let the demand vector be denoted by $\textbf{d} = (d_1,d_2,\dots,d_K)$. To derive an upper bound on the required transmission rate, we focus our attention on the worst case scenario i.e. each user requests for distinct files. 
	The delivery steps are presented as an algorithm in {\bf Algorithm 1}.
\begin{algorithm}
		\caption{Algorithm for Delivery}
		\begin{algorithmic}[1]
			\For {$u=1$ to $u=\binom{r}{z}$}
			\State Choose a set of $z$ out of $r$ parallel classes
			\State which is different from the sets chosen before. 
			\State Let this set be
                        $$\mathcal{P}_1 = \{C_{1,1},C_{1,2},\dots,C_{1,i_1},\dots,C_{1,b_r}\},$$
			$$\mathcal{P}_2 = \{C_{2,1},C_{2,2},\dots,C_{2,i_2},\dots,C_{2,b_r}\},$$
			$$\vdots$$
			$$\mathcal{P}_z = \{C_{z,1},C_{z,2},\dots,C_{z,i_z},\dots,C_{z,b_r}\}.$$
			\For{$v=1$ to $v=\binom{b_r}{2}^z$}
			\State Choose a pair of blocks from each of the $z$
			\State parallel classes $\mathcal{P}_1,\mathcal{P}_2,\dots,\mathcal{P}_z$.  This set of $2z$ 
			\State blocks must be different from the ones chosen   
                        \State before. Let the chosen set be
			$$\{C_{1,i_1},C_{1,j_1},C_{2,i_2},C_{2,j_2},\dots,C_{z,i_z}C_{z,j_z}\}$$
			\State where, $i_s, j_s \in [b_r] \text{ and } j_s \neq i_s, \forall s \in [z]$.
			\State There are $2^z$ users corresponding to the $2z$ blocks 
			\State chosen above. Denoting this set of user indices by
			\State $\mathcal{X,}$ we have 
			$$\;\;\;\mathcal{X}=\{(C_1',C_2',\dots,C_z') : C_s' \in \{C_{s,i_s},C_{s,j_s}\}\}$$
			\State \textbf{Calculate:} Calling the user connected to the 
                         \State set of caches $\{C_{1,a_1},C_{2,a_2},\dots,C_{z,a_z}\}$, 
                         \State where $a_k \in \{i_k,j_k \},$  $k=1,2, \cdots,z,$ to be 
                         \State the  $m^{th}$ user calculate the set $f_m$ as 
			$$f_m = C_{1,e_1}\cap C_{2,e_2}\cap\dots\cap C_{z,e_{z}}$$
                        \State where $e_k=\{i_k,j_k \}\setminus a_k.$ We have $|f_m|=\mu_{z}$. Let
			$$f_m := \{y_{m,1},y_{m,2},\dots,y_{m,\mu_{z}}\}. $$
			\State Calculate $f_m$ as above for all the $2^z$ users in $\mathcal{X}.$
                        \State \textbf{Transmit:} Now do the following $\mu_z$ transmissions
			 $$\underset {m \in \mathcal{X}} \oplus W_{d_{m},y_{m,s}},\;\forall s \in [\mu_{z}]$$
			\State Note that there are $\mu_{z}$ transmissions for  $\mathcal{X.}$
			\EndFor
			\EndFor
		\end{algorithmic}
	\end{algorithm}

The proof of correctness of the {\bf Algorithm 1} is given in the Appendix.  We have 
\begin{thm}
\label{thm1}
For $N$ files and $K$ users each with access to $z$ caches of size $M$ in the considered caching system, if it $N\geq K$ and for the distinct demands by the users, the proposed scheme achieves the rate $R$ given by
                $$R = \frac{\mu_z\binom{b_r}{2}^z\binom{r}{z}}{v}$$
\end{thm}
\begin{IEEEproof}
                The first \textbf{for} loop of the delivery algorithm runs $\binom{r}{z}$ times. The second \textbf{for} loop of the delivery algorithm runs $\binom{b_r}{2}^z$ times. The transmit step of the delivery algorithm runs $\mu_z$ times.
                So we see that totally there are $\mu_z\binom{b_r}{2}^z\binom{r}{z}$ transmissions and subpacketization level is $v$. Hence, the result from the definition of rate.
\end{IEEEproof}
\begin{lem}
\label{lem4}
Number of users benefited in each transmission, known in the literature as coding gain and denoted by $g,$ is given by
                $$ g = 2^{z}$$
\end{lem}
\begin{IEEEproof}
                From second \textbf{for} of the delivery algorithm it can be observed that, there are totally $2^z$ users benefited from a transmission. So the coding gain by definition is $2^z$.
\end{IEEEproof}
\section{Performance Comparison}
\label{sec4}

For both the MaN and the SPE schemes the number of users and the number of caches are same. Whereas for the schemes proposed in this paper the number of users $K$ and the number of caches $b$ need not be same. So, when we compare our scheme with the MaN or the SPE scheme we will compare taking  the number of caches being equal and also the number of caches a user can access also being same. 
\subsection{Comparison of our schemes obtained from CRDs from Affine Planes}
In this subsection we focus on the schemes obtained using  the resolvable designs from affine planes \cite{Stinson} which are CRDs.  We will compare the resulting schemes with the MaN scheme. Such CRDs exists for all $n$ where $n$ is a prime or a power of a prime number. For an $n,$ the CRD resulting from an affine plane has the number of points  $v=n^2,$ the number of points in a block $k=n,$ the number of blocks $b=n(n+1),$ the number of parallel classes $r=n+1.$   It is known that any two blocks drawn from different parallel classes always intersect at exactly one point \cite{Stinson} i.e., $z=2$ and $\mu_z=1.$

For the multi-access coded caching scheme from the CRD with parameter $n,$ with $N$ files we have $K=(b_r)^z\binom{r}{z}=\frac{n^3(n+1)}{2}$ users, $b=n(n+1)$ caches each having a cache size of $\frac{M}{N}=\frac{k}{v}=\frac{1}{n}$ and  $\frac{M'}{N} = \frac{(2n-1)}{n^2}.$ 
\subsubsection{Comparison with the MaN Scheme}
\begin{figure}
    \begin{center}
    \includegraphics[width=9cm,height=6cm]{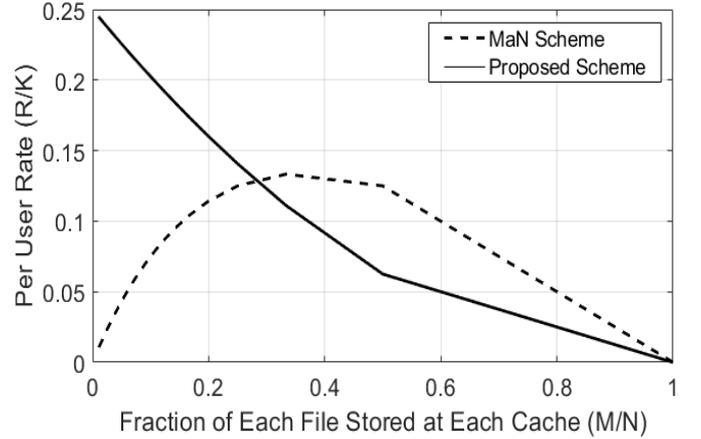}
    \caption {Performance Analysis between MaN and Proposed scheme for the class of cross resolvable design derived from affine planes for the case $z=2$,  where n is a prime or prime power.}
\label{fig2}
    \end{center}
\end{figure}

Since $\frac{bM}{N}=\frac{b}{n} = (n+1)$ is a integer, we have the corresponding MaN scheme with $N$ files, $n(n+1)$ number of users and each user has $\frac{1}{n}$ fraction of each file stored at its corresponding cache. The two schemes have been compared by keeping the number of caches and fractions of each file at each cache equal in Table \ref{tab1}. It is seen that our scheme performs better than the MaN scheme in terms of the number of users supported and the subpacketization level at the cost of increased rate and decreased gain. Since the fraction of each file at each cache $\frac{M}{N}$ is the same in both the schemes, in Fig.\ref{fig2} the per user rate $\frac{R}{K}$ is plotted against $\frac{M}{N}$ from which it is clear that our proposed scheme performs better than the MaN scheme for large cache sizes (approximately 0.3 onwards). For smaller cache sizes the MaN scheme performs better.  
\begin{table}
\caption{Comparison between the MaN and proposed scheme for the class of CRDs from affine planes where $n$ is a prime or prime power.}
  \begin{center}
  \renewcommand{\arraystretch}{2}
    \begin{tabular}{|c|c|c|}
    \hline
      \textbf{Parameters} &\textbf{MaN Scheme}  &  \textbf{Proposed Scheme}\\\hline\hline
      Number of Caches & $n(n+1)$ &  $n(n+1)$\\\hline
      \makecell{Fraction of each file \\at each cache $\left(\frac{M}{N}\right)$} & $\frac{1}{n}$ &  $\frac{1}{n}$\\\hline
      Number of Users $(K)$ & $n(n+1)$   &$\frac{n^3(n+1)}{2}$\\\hline
      Subpacketization level  & $\binom{n(n+1)}{n+1}$ & $n^2$\\[.2cm]\hline
      Rate $(R)$  & $\frac{(n+1)(n-1)}{n+2}$ & $\frac{n(n+1)(n-1)^2}{8}$\\[.2cm]\hline
      Gain $(g)$ & $n+2$  & $4$\\\hline
    \end{tabular}
  \end{center}
\label{tab1}
\end{table}

\subsubsection{Comparison with the SPE scheme}
\begin{figure}
    \begin{center}
    \includegraphics[width=7.0cm, height=7.0cm]{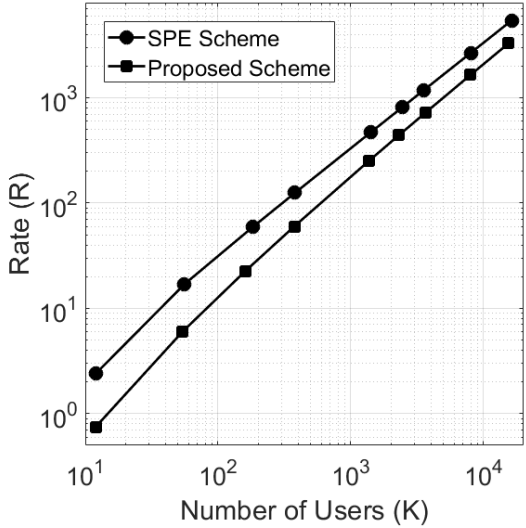}
    \caption {Rate $R$ for the schemes in Table \ref{tab2}}
\label{fig3}
    \end{center}
\end{figure}

\begin{figure}
      \begin{center}
       \includegraphics[width=9cm,height=6cm]{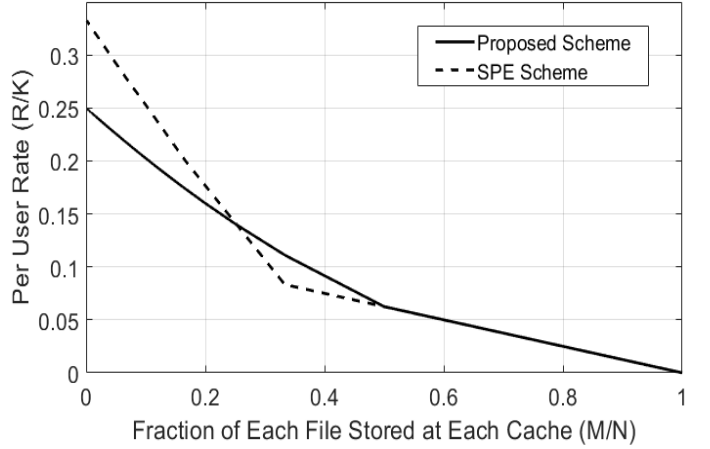}
       \caption {Per user rate for the codes in Table \ref{tab2}}
\label{fig4}       
\end{center}
\end{figure}

For the subpacketization value to be an integer in \cite{SPE}, given by $\frac{K(K-2z+2)}{4}$, where $K$ is the number of users and $z$ is the number of caches a user has access to, we consider the SPE scheme with $N$ files, $K = n(n+1)$ users, $\frac{M}{N} = \frac{2}{n(n+1)}$ fraction of each file stored at each cache and each user has access to exactly $z = 2$ caches, which gives a subpacketization level equal to $\frac{n(n-1)(n+1)(n+2)}{4}$. Since $\frac{b(b-2z+2)}{4} = \frac{n(n-1)(n+1)(n+2)}{4}$ is an integer, we have the comparable SPE scheme with $N$ files, $n(n+1)$ number of users. The comparison is given in Table \ref{tab2} and also shown in Fig.\ref{fig3}. The rate expression in \cite{SPE} is complicated (given in Theorem $1$, page $2$ of \cite{SPE}). So for comparison with the proposed scheme we plot the rate versus the number of users for the two schemes in Fig. \ref{fig3}. It is seen that our scheme outperforms the SPE scheme in the number of users allowed, subpacketization level, rate as well as in gain at the cost of increase in the fraction of each file stored in each cache. In Fig.\ref{fig4} we plot the per user rate $\frac{R}{K}$ against the fraction of each file stored at each cache $\frac{M}{N}$ and see that our scheme performs better than the SPE scheme for small cache sizes and matches the performance at large cache sizes. 
\begin{table}
\caption{Comparison between SPE and Proposed scheme for the class of CRDs from affine planes where $n$ is a prime or prime power.}
  \begin{center}
  \renewcommand{\arraystretch}{2.5}
    \begin{tabular}{|c|c|c|}
    \hline
      \textbf{Parameters} &\textbf{SPE Scheme}  &  \textbf{Proposed Scheme}\\\hline\hline
      Number of Caches & $n(n+1)$ &  $n(n+1)$\\\hline
      \makecell{Number of Caches a \\user has access to $(z)$} & $2$ &  $2$\\\hline
      Number of Users $(K)$ & $n(n+1)$   &$\frac{n^3(n+1)}{2}$\\\hline
      \makecell{Fraction of each file \\at each cache $\left(\frac{M}{N}\right)$} & $\frac{2}{n(n+1)}$ &  $\frac{1}{n}$\\\hline
      \makecell{Fraction of each file each\\user has access to } & $\frac{4}{n(n+1)}$ &  $\frac{2n-1}{n^2}$\\\hline
      Subpacketization level & $\frac{n(n-1)(n+1)(n+2)}{4}$ & $n^2$\\[.2cm]\hline
      Rate (R) & See Fig. \ref{fig3} & See Fig.\ref{fig3} \\ \hline 
      Gain $(g)$ & between $3$ and $4$  & $4$\\\hline
    \end{tabular}
  \end{center}
\label{tab2}
\end{table}
\subsection{Comparison of our schemes obtained from CRDs from Affine Resolvable BIBDs}
In this subsection we focus on the schemes obtained using the resolvable designs from affine resolvable balanced incomplete block designs (BIBDs) \cite{Stinson} which are CRDs. We will compare the resulting schemes with the MaN scheme. One such infinite class of affine resolvable BIBDs is derived from affine geometry. Such CRDs exists for all $q$ and $m,$ where $q$ is a prime or a power of a prime number and $m\geq2$. For an $q$ and $m,$ the CRD resulting from an affine resolvable BIBD has the number of points $v=q^m,$ the number of points in a block $k=q^{m-1},$ the number of blocks $b=\frac{q(q^m-1)}{q-1},$ the number of parallel classes $r=\frac{q^m-1}{q-1}.$ It is known that any two blocks drawn from different parallel classes always intersect at exactly $\frac{k^2}{v}$ points \cite{Stinson} i.e., $z=2$ and $\mu_z=q^{m-2}.$

The special case of $m=2$ is the class of CRDs from affine planes discussed in the previous subsection.

For the multi-access coded caching scheme from the CRD with parameter $q$ and $m,$ with $N$ files we have $K=(b_r)^z\binom{r}{z}=\frac{q^3(q^m-1)(q^{m-1}-1)}{2(q-1)^2}$ users, $b=\frac{q(q^m-1)}{q-1}$ caches each having a cache size of $\frac{M}{N}=\frac{k}{v}=\frac{1}{q}$ and $\frac{M'}{N} = \frac{(2q-1)}{q^2}.$

Since $\frac{bM}{N}=\frac{b}{q} = \frac{(q^m-1)}{q-1}$ is a integer, we have the corresponding MaN scheme with $N$ files, $\frac{q(q^m-1)}{q-1}$ number of users and each user has $\frac{1}{q}$ fraction of each file stored at its corresponding cache. The two schemes have been compared by keeping the number of caches and fractions of each file at each cache equal in Table \ref{tab3}. It is seen that our scheme performs better than the MaN scheme in terms of the number of users supported and the subpacketization level at the cost of increased rate and decreased gain.

\begin{table}[H]
\caption{Comparison between MaN and Proposed scheme for the class of cross resolvable design derived from affine geometry for the case $z=2$, where $q$ is a prime or prime power and $m\geq2$.}

\begin{center}
\renewcommand{\arraystretch}{2.5}
\begin{tabular}{|c|c|c|}
\hline
\textbf{Parameters} &\textbf{MaN Scheme} & \textbf{Proposed Scheme}\\\hline\hline
\makecell{Number of\\ Caches} & $\dfrac{q(q^m-1)}{q-1}$ & $\dfrac{q(q^m-1)}{q-1}$\\[.2cm]\hline
\makecell{Fraction of each\\ file at each\\ cache $\left(\frac{M}{N}\right)$} & $\dfrac{1}{q}$ & $\dfrac{1}{q}$\\\hline
\makecell{Number of\\ Users $(K)$} & $\dfrac{q(q^m-1)}{q-1}$ &$\dfrac{q^3(q^m-1)(q^{m-1}-1)}{2(q-1)^2}$\\[.2cm]\hline
\makecell{Subpackelization\\ level $(F)$} & $\dbinom{q(q^m-1)/q-1}{(q^m-1)/q-1}$ & $q^m$\\[.2cm]\hline
Rate $(R)$ & $\dfrac{(q^m-1)(q-1)}{q^m+q-2}$ & $\dfrac{q(q^m-1)(q^{m-1}-1)}{8}$\\[.2cm]\hline
\makecell{Rate per\\ user $\left(\frac{R}{K}\right)$}&$\dfrac{(q-1)^2}{q(q^m+q-2)}$&$\dfrac{(q-1)^2}{4q^2}$\\[.2cm]\hline
Gain $(g)$ & $\dfrac{q^m-1}{q-1}+1$ & $4$\\[.2cm]\hline
\end{tabular}
\end{center}
\label{tab3}
\end{table}

Another such infinite class of affine resolvable BIBDs is derived from Hadamard matrices. Such CRDs exists for all $m,$ if there exists a Hadamard matrix of order $4m$. For an $m,$ the CRD resulting from an affine resolvable BIBD has the number of points $v=4m,$ the number of points in a block $k=2m,$ the number of blocks $b=2(4m-1),$ the number of parallel classes $r=4m-1.$ It is known that any two blocks drawn from different parallel classes always intersect at exactly $\frac{k^2}{v}$ points \cite{Stinson} i.e., $z=2$ and $\mu_z=m.$

For the multi-access coded caching scheme from the CRD with parameter $q$ and $m,$ with $N$ files we have $K=(b_r)^z\binom{r}{z}=4(2m-1)(4m-1)$ users, $b=2(4m-1)$ caches each having a cache size of $\frac{M}{N}=\frac{k}{v}=\frac{1}{2}$ and $\frac{M'}{N} = \frac{3}{4}.$ 

Since $\frac{bM}{N}=\frac{b}{2} = 4m-1$ is a integer, we have the corresponding MaN scheme with $N$ files, $2(4m-1)$ number of users and each user has $\frac{1}{2}$ fraction of each file stored at its corresponding cache. The two schemes have been compared by keeping the number of caches and fractions of each file at each cache equal in Table \ref{tab4}. It is seen that our scheme performs better than the MaN scheme in terms of the number of users supported and the subpacketization level at the cost of increased rate and decreased gain.

\begin{table}[H]
\caption{Comparison between MaN and Proposed scheme for the class of cross resolvable design derived from Hadamard matrices for the case $z=2$.}
\begin{center}
\renewcommand{\arraystretch}{2.5}
\begin{tabular}{|c|c|c|}
\hline
\textbf{Parameters} &\textbf{MaN Scheme} & \textbf{Proposed Scheme}\\\hline\hline
\makecell{Number of Caches} & $2(4m-1)$ & $2(4m-1)$\\\hline
\makecell{Fraction of each file\\ at each cache $\left(\frac{M}{N}\right)$} & $\frac{1}{2}$ & $\frac{1}{2}$\\\hline
\makecell{Number of Users $(K)$} & $2(4m-1)$ &$4(2m-1)(4m-1)$\\\hline
\makecell{Subpackelization level $(F)$} & $\binom{2(4m-1)}{4m-1}$ & $4m$\\\hline
Rate $(R)$ & $\frac{(4m-1)}{4m}$ & $\frac{(2m-1)(4m-1)}{4}$\\\hline
Rate per user $\left(\frac{R}{K}\right)$ &$\frac{1}{8m}$&$\frac{1}{16}$\\\hline
Gain $(g)$ & $4m$ & $4$\\\hline
\end{tabular}
\end{center}
\label{tab4}
\end{table}

Affine resolvable BIBDs derived either from affine geometry or from Hadamard matrices has cross intersection number (i.e. $\mu_2$) equal $\frac{k^2}{v}$. When this cross intersection number is minimal i.e $\mu_2 = 1$, we see that our scheme performs better than MaN scheme. One such design is affine resolvable BIBDs derived from affine geometry for $m=2.$

\begin{figure}[H]
\begin{center}
\includegraphics[width=8.8cm,height=8cm]{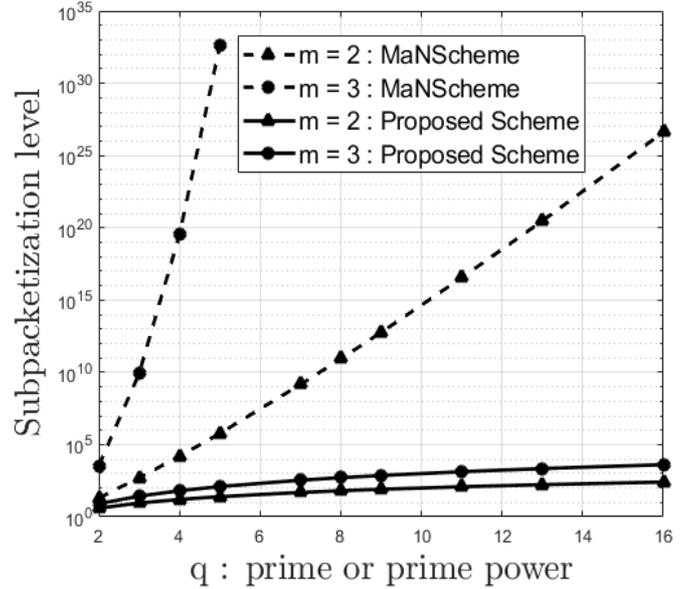}
\caption {Comparing subpacketization level for $m=2$ and $m=3$ between MaN and Proposed scheme for the class of resolvable designs derived from affine resolvable BIBD's , where q is a prime or prime power.}
\label{fig5}
\end{center}
 \end{figure}
\begin{figure}
\begin{center}
\includegraphics[width=9cm,height=7.5cm]{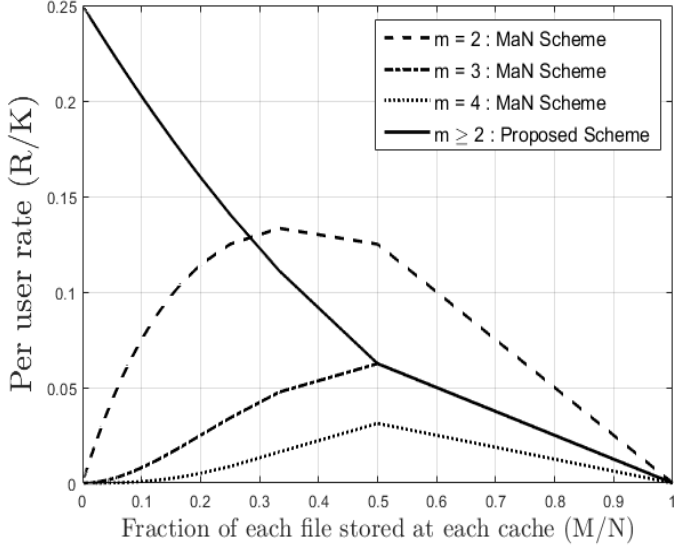}
\caption {Performance Analysis between MaN and Proposed scheme for the class of cross resolvable design derived from affine resolvable BIBD's for the case $z=2$,  where q is a prime or prime power and $m \geq 2$.}
\label{fig6}
\end{center}
\end{figure}
Fig. \ref{fig5} shows the subpacketization $v$ verses the parameter $q.$ 
The plot in Fig. \ref{fig6} shows the per user rate $\frac{R}{K}$ verses the fraction of each file stored at each cache $\frac{M}{N}$. Since $\frac{M}{N} = \frac{1}{q}$ and $\frac{R}{K} = \frac{(q-1)^2}{(q)(q^m+q-2)}$ in case of MaN scheme, $\frac{R}{K} = \frac{(q-1)^2}{4q^2}$ in case of Proposed scheme is a function of $q$ ,  we plot $\frac{R}{K}$ vs $\frac{M}{N}$ keeping $m$ constant for different values of $q$, where $q$ is a prime or prime power number. The points between any two  achievable $(\frac{M}{N}, \frac{R}{K})$  pairs through this scheme, can be obtained as well through memory sharing. Notice that Fig. \ref{fig2} appears again in Fig. \ref{fig6}.
\subsection{Two examples outperforming the MaN scheme}
In this subsection, we present two instances of our schemes using CRDs (not from affine planes) which outperform the MaN scheme in all aspects, namely in rate, gain as well as in subpacketization level simultaneously. The first instance is our scheme obtained using the CRD given in Example \ref{exmp3} and the second one is that obtained from the CRD given in Example \ref{exmp4}. The performances of these two schemes in comparison with the comparable  MaN schemes is presented in Table \ref{tab34}.

\begin{figure}
       \begin{center}
       \includegraphics[width=9cm,height=5cm]{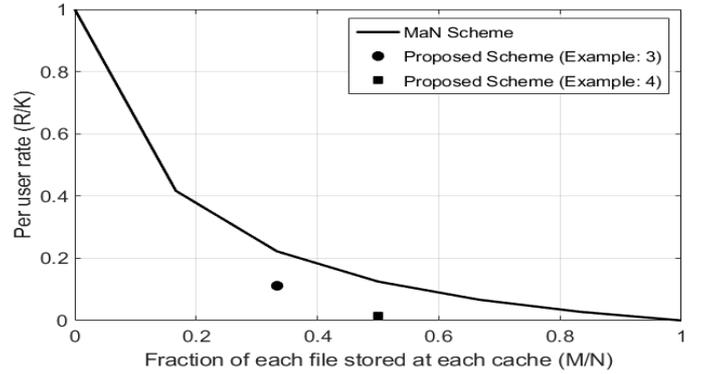}
        \caption {Pictorial representation of Table \ref{tab34}}
\label{fig7}
        \end{center}
\end{figure}

\begin{table}
\caption{Comparison between MaN and proposed scheme}
  \begin{center}
  \renewcommand{\arraystretch}{2}
    \begin{tabular}{|c||c|c||c|c|}
    \hline
    \multirow{2}{*}{\textbf{Parameters}}&\multicolumn{2}{c||}{\textbf{CRD of Example 3}} & \multicolumn{2}{c|}{\textbf{CRD of Example 4}}\\
    \cline{2-5}
     &\textbf{\makecell{MaN\\Scheme}}&\textbf{\makecell{Proposed\\Scheme}}&\textbf{\makecell{MaN\\Scheme}}&\textbf{\makecell{Proposed\\Scheme}}\\
      \hline
      \hline
      \makecell{Number of\\ Caches $(b)$}& $6$ & $6$& $6$ & $6$\\\hline
      \makecell{Number of \\Caches a user\\ has access to} & $1$ & $2$& $1$ & $3$\\\hline
      \makecell{Number of\\ Users $(K)$} & $6$ & $9$& $6$ & $8$\\\hline
      \makecell{Subpacketization\\level $(F)$} & $15$ &  $9$& $20$ &  $8$\\\hline
      \makecell{Fraction of each\\ file at each\\ cache $\left(\frac{M}{N}\right)$} & $\frac{1}{3}$ &$\frac{1}{3}$& $\frac{1}{2}$ &  $\frac{1}{2}$\\\hline
      \makecell{Fraction of each\\ file  each user\\ has access to}& $\frac{1}{3}$  &  $\frac{5}{9}$& $\frac{1}{2}$  &  $\frac{7}{8}$\\ \hline
      Rate $(R)$  & $\frac{4}{3}$  & $1$& $\frac{3}{4}$  & $\frac{1}{8}$\\\hline
      Gain $(g)$ & $3$ &  $4$& $4$ &  $8$\\
    \hline
    \end{tabular}
  \end{center}
\label{tab34}
\end{table}
The performance improvement of our scheme for these two instances is shown pictorially in Fig. \ref{fig7}
\subsection{Comparison with SPE scheme}
In this subsection, we show some examples (not necessarily from affine planes) for the comparison between two schemes.
\begin{exmp}
\label{exmp7}\\
Consider the resolvable design with parameters $v=8,\;b=8,\;r=4,\;k=4$ and $\mu_2=2$ specified as follows.
\begin{align*}
        X =&\; \{1, 2, 3, 4, 5, 6, 7, 8\}, \text{and} \\
        \mathcal{A} = &\;\{\{1, 2, 3, 4\},\{ 5, 6, 7, 8\},\{1, 2, 5, 6\},\{ 1, 3, 5, 7\},\\
        &\{2, 4, 6, 8\},\{ 3, 4, 7, 8\},\{1, 4, 5, 8\},\{ 2, 3, 6, 7\}\}
        \end{align*}
        The parallel classes are $\mathcal{P}_1 =\; \{\{1, 2, 3, 4\},\{ 5, 6, 7, 8\}\},$ $\mathcal{P}_2 =\; \{\{1, 2, 5, 6\},\{ 3, 4, 7, 8\}\},$ $\mathcal{P}_3 =\; \{\{2, 4, 6, 8\},\{ 1, 3, 5, 7\}\} $ and $\mathcal{P}_4 =\; \{\{1, 4, 5, 8\},\{ 2, 3, 6, 7\}\}.$
\end{exmp}
\begin{table}
\caption{Comparison between SPE and proposed scheme}
  \begin{center}
  \renewcommand{\arraystretch}{2}
    \begin{tabular}{|c||c|c||c|c|}
    \hline
    \multirow{2}{*}{\textbf{Parameters}}&\multicolumn{2}{c||}{\textbf{CRD of Example 5}} & \multicolumn{2}{c|}{\textbf{CRD of Example 4}}\\
    \cline{2-5}
     &\textbf{\makecell{SPE\\Scheme}}&\textbf{\makecell{Proposed\\Scheme}}&\textbf{\makecell{SPE\\Scheme}}&\textbf{\makecell{Proposed\\Scheme}}\\
      \hline
      \hline
      \makecell{Number of\\ Caches $(b)$}& $8$ & $8$& $6$ & $6$\\\hline
      \makecell{Number of \\Caches a user\\ has access to} & $2$ & $2$& $2$ & $2$\\\hline
      \makecell{Number of\\ Users $(K)$} & $8$ & $24$& $6$ & $12$\\\hline
      \makecell{Subpacketization\\level $(F)$} & $12$ &  $8$& $6$ &  $8$\\\hline
      \makecell{Fraction of each\\ file at each\\ cache $\left(\frac{M}{N}\right)$} & $\frac{1}{4}$ &$\frac{1}{2}$& $\frac{1}{3}$ &  $\frac{1}{2}$\\\hline
      \makecell{Fraction of each\\ file  each user\\ has access to}& $\frac{1}{2}$  &  $\frac{3}{4}$& $\frac{2}{3}$  &  $\frac{3}{4}$\\ \hline
      Rate $(R)$  & $\frac{7}{6}$  & $\frac{3}{2}$& $\frac{1}{2}$  & $\frac{3}{4}$\\\hline
      Gain $(g)$ & $\frac{25}{7}$ &  $4$& $4$ &  $4$\\
    \hline
    \end{tabular}
  \end{center}
\label{tab56}
\end{table}
The comparison between the SPE scheme and proposed scheme in Table \ref{tab56} shows that more users can be supported in a multi-access setup, and certain choices of cross resolvable designs can yield better subpacketization levels and even better gains than the comparable SPE scheme at the cost of increased storage in each cache.
\section{Schemes from Resolvable Designs  ($z=1$)}
\label{sec5}
So far, we have considered the multi-access coded caching problems from CRDs where a user is associated with   $z,\;z\in\{2,3,\dots,r\}$ distinct caches such that these $z$ caches correspond to $z$ blocks from distinct parallel classes. In this section we describe multi-access coded caching schemes from resolvable designs that are not CRD. For such resolvable designs we define $z=1$. We see that this case is nothing but the case where  each user is associated with its own dedicated cache, i.e., the number of users is equal to the number of blocks. 

For a given resolvable design $(X, \mathcal{A})$ with $v$ points, $r$ parallel classes, $b$ blocks of size $k$ each, $b_{r}$ blocks in each parallel class, now we have the coded caching problem with  $N$ files in server database, $b$ be the number of caches, $K = {r \choose z} b_{r}^{z} = rb_r= b$ users, $\frac{M}{N} = \frac{k}{v}$ fraction of each file at each cache and subpacketization level $v$.
\subsection{Placement and Delivery Phase}
The placement phase remains the same as described in Section \ref{sec2} for schemes from CRDs  and the Delivery steps are same as in  as {\bf Algorithm 1} with taking $\mu_1=k.$ It can be checked that the proof of correctness of {\bf Algorithm 1} holds 
for the case included with $z=1$ and $\mu_1=k.$ 
Now Theorem \ref{thm1} and Lemma \ref{lem4} can be restated for the case $z=1$ as follows with the proofs remaining valid.
\begin{thm}
\label{thm2} 
For multi-access coded caching schemes from resolvable designs with $z=1,$  $N$ files and $K$ users each with access to one cache of size $M$ in the considered caching system, with $N\geq K$, the proposed scheme achieves the worst case rate $R$ with distinct demands given by $R = \frac{rk\binom{b_r}{2}}{v}$ and  the number of users benefited in each transmission i.e gain, is 2.
\end{thm}
%
\subsection{Performance Comparison: MaN scheme vs schemes from resolvable designs from affine planes with $z=1$}
Consider the coded caching schemes from  the resolvable design derived from the affine plane treating them with $z=1.$ We have  $b = K = n(n+1)$ and $\frac{M}{N} = \frac{1}{n}$. Since $\frac{KM}{N} = n+1$ is an integer, we have the corresponding MaN scheme with $N$ files, $n(n+1)$ number of users and each user having $\frac{1}{n}$ fraction of each file stored at it's corresponding  cache. Comparison of proposed scheme with MaN scheme is given in Table \ref{tab5}.
\begin{table}[H]
\caption{Comparison between MaN  and Proposed scheme for the class of resolvable designs derived from affine planes with $z=1$.}
  \begin{center}
  \renewcommand{\arraystretch}{2}
    \begin{tabular}{|c|c|c|}
    \hline
      \textbf{Parameters} &\textbf{MaN Scheme}  &  \textbf{Proposed Scheme}\\\hline\hline
      Number of Caches & $n(n+1)$ &  $n(n+1)$\\\hline
      \makecell{Fraction of each file \\at each cache $\left(\frac{M}{N}\right)$} & $\frac{1}{n}$ &  $\frac{1}{n}$\\\hline
      Number of Users $(K)$ & $n(n+1)$   &$n(n+1)$\\\hline
      Subpacketization level $(F)$ & $\binom{n(n+1)}{n+1}$ & $n^2$\\[.2cm]\hline
      Rate $(R)$  & $\frac{(n+1)(n-1)}{n+2}$ & $\frac{(n+1)(n-1)}{2}$\\[.2cm]\hline
      Gain $(g)$ & $n+2$  & $2$\\\hline
    \end{tabular}
  \end{center}
  \label{tab5}
\end{table}
\subsection{Comparison of proposed scheme for different values of $z$}
In this subsection, we analyze the proposed scheme for different values of $z\in \{2,3,\dots,r\}$. We will analyze a CRD for different values of $z$ with respect to rate per user $\left(\frac{R}{K}\right)$.
\begin{equation}
\label{z2tor}
    \frac{\left(\frac{R}{K}\right)_{z}}{\left(\frac{R}{K}\right)_{z-1}} = 
    \frac{1}{2}\left(1-\frac{k}{v}\right),~~~~ z \in \{3,4,\dots,r\}
\end{equation}
\begin{equation}
\label{z2to2}    
\frac{\left(\frac{R}{K}\right)_{z}}{\left(\frac{R}{K}\right)_{z-1}} = 
    \frac{\mu_2}{2k}\left(\frac{v}{k}-1\right),~~~~  z = 2
\end{equation}
\begin{exmp}
\label{exmp8}
Consider the resolvable design  with parameters $v=27,\;b=9,\;r=3,\;k=9,\;\mu_2 = 3$ and $\mu_3 = 1$.
In Table \ref{tab8}, we compare the  resolvable design for different values of $z$.
\begin{table}[H]
\caption{Comparison for different values of $z$ for a resolvable design given in Example \ref{exmp8}}
  \begin{center}
  \renewcommand{\arraystretch}{3}
    \begin{tabular}{|c|c|c|c|}
    \hline
      \textbf{Parameters}&\textbf{z = 1}&\textbf{z = 2}&\textbf{z = 3}\\
      \hline
      \hline
      \makecell{Number of Caches $(b)$} & $9$ & $9$& $9$\\\hline
      \makecell{Subpacketization level $(F)$} & $27$ & $27$ & $27$\\\hline
      \makecell{Number of Users $(K)$} & $9$ & $27$& $27$ \\\hline
      \makecell{Fraction of each file\\ at each cache $\left(\frac{M}{N}\right)$ }& $\frac{1}{3}$&  $\frac{1}{3}$& $\frac{1}{3}$ \\\hline
      \makecell{Fraction of each file\\  each user has access to $\left(\frac{M'}{N}\right)$}& $\frac{1}{3}$&  $\frac{5}{9}$& $\frac{19}{27}$\\\hline
      Rate $(R)$  & $3$  & $3$& $1$\\\hline
      Rate  per user $\left(\frac{R}{K}\right)$ &$\frac{1}{3}$&$\frac{1}{9}$&$\frac{1}{27}$\\\hline
      Gain $(g)$ & $2$ &  $4$& $8$\\
    \hline
    \end{tabular}
  \end{center}
  \label{tab8}
\end{table}
\end{exmp}
\begin{exmp}
\label{exmp9}
Consider the resolvable design  with parameters $v=16,\;b=8,\;r=4$ and $\;k=8$ as given below
 \begin{align*}
    X = &\;\{1, 2, 3, 4, 5, 6, 7, 8, 9, 10, 11, 12, 13, 14, 15, 16\}, \text{ and} \\
    \mathcal{A} = & \;\{\{1, 2, 3, 4, 5, 6, 7, 8\},\{9, 10, 11, 12, 13, 14, 15, 16\}\\
    &\;\{1, 2, 3, 4, 9, 10, 11, 12\},\{5, 6, 7, 8, 13, 14, 15, 16\}\\
    &\;\{1, 2, 5, 6, 9, 10, 13, 14\},\{3, 4, 7, 8, 11, 12, 15, 16\}\\
    &\;\{1, 3, 5, 7, 9, 11, 13, 15\},\{2, 4, 6, 8, 10, 12, 14, 16\}\}\\
    \end{align*}
    The parallel classes are
    \begin{align*}
    \mathcal{P}_1 =&\; \{\{1, 2, 3, 4, 5, 6, 7, 8\},\{9, 10, 11, 12, 13, 14, 15, 16\}\},\\
    \mathcal{P}_2 =&\; \{\{1, 2, 3, 4, 9, 10, 11, 12\},\{5, 6, 7, 8, 13, 14, 15, 16\}\},\\
    \mathcal{P}_3 =&\; \{\{1, 2, 5, 6, 9, 10, 13, 14\},\{3, 4, 7, 8, 11, 12, 15, 16\}\},\\
    \mathcal{P}_4 =&\; \{\{1, 3, 5, 7, 9, 11, 13, 15\},\{2, 4, 6, 8, 10, 12, 14, 16\}\}
    \end{align*}

It can be easily observed that $\mu_2 = 4, \mu_3 = 2$ and $\mu_4 = 1$. In Table \ref{tab9}, we compare the  resolvable design for different values of $z$.
\begin{table}[H]
\caption{Comparison for different values of $z$ for a resolvable design given in Example \ref{exmp9}}
  \begin{center}
  \renewcommand{\arraystretch}{3}
    \begin{tabular}{|c|c|c|c|c|}
    \hline
      \textbf{Parameters}&\textbf{z = 1}&\textbf{z = 2}&\textbf{z = 3}&\textbf{z = 4}\\
      \hline
      \hline
      Number of Caches $(b)$ & $8$ & $8$& $8$ & $8$\\\hline
      Subpacketization level $(F)$ & $16$ & $16$& $16$ & $16$\\\hline
      Number of Users $(K)$ & $8$ & $24$& $32$ & $16$\\\hline
      \makecell{Fraction of each file\\ at each cache $\left(\frac{M}{N}\right)$ }& $\frac{1}{2}$&  $\frac{1}{2}$& $\frac{1}{2}$ &  $\frac{1}{2}$\\\hline
      \makecell{Fraction of each file \\ each user has access to $\left(\frac{M'}{N}\right)$}& $\frac{1}{2}$&  $\frac{3}{4}$& $\frac{7}{8}$ &  $\frac{15}{16}$\\\hline
      Rate $(R)$  & $2$  & $\frac{3}{2}$& $\frac{1}{2}$  & $\frac{1}{16}$\\\hline
      Rate  per user $\left(\frac{R}{K}\right)$ &$\frac{1}{4}$&$\frac{1}{16}$&$\frac{1}{64}$&$\frac{1}{256}$\\\hline
      Gain $(g)$ & $2$ &  $4$& $8$ &  $16$\\
    \hline
    \end{tabular}
  \end{center}
  \label{tab9}
\end{table}
\end{exmp}

From Table \ref{tab8}, we can see that though one would expect more $z$, to give more improvement in overall rate, this need not be the case. For $z = 1$ and for $z = 2$ the overall rate remains same, though gain improves from $2$ to $4$. This is because the number of users supported improved from $9$ to $27$. Comparing in terms of rate per user, in both the examples we see that as $z$ increases, rate per user decreases. At the same time, it need not be the case that increasing value of $z$ , yields increase in number of users that can be supported. This is evident from Table \ref{tab9}. We see a drop in number of users supported from $32$ to $16$ when $z$ changes from $z = 3$ to $z = 4$. 
\subsection{Subpacketization}

\begin{lem}
For a given resolvable design with parameters $v,\;b,\;r,\;k$ and for the chosen value of $z$, with $K$ being the number of users supported, the subpacketization level $v$ is given by
\begin{equation}
\label{subpaklevel}
v = k\left( \frac{K}{\binom{r}{z}}   \right)^{1/z}.
\end{equation}
\end{lem}
\begin{IEEEproof}
 Since for a given resolvable design and for a chosen value of $z$, number of users $K$ is equal to 
 $$ K = \binom{r}{z}(b_r)^z = \binom{r}{z}\left(\frac{v}{k}\right)^z.$$
Rearranging the terms of the equation above gives (\ref{subpaklevel}).
\end{IEEEproof}

From the above lemma, it is not hard to see that the subpacketization level $v$ grows slower than the number of users $K$ i.e. the subpacketization level growth is  sublinear with $K$ for the different classes of cross resolvable designs discussed in this paper.
\section{Discussion}
\label{sec6}
We have identified a special class of resolvable designs called cross resolvable designs which lead to multi-access coded caching schemes. While combinatorial designs have been used in the literature for coded caching problems ours is the first work to use them for multi-access coded caching. Our results indicate that using CRDs in multi-access setups can help attain gains beyond cache redundancy at low subpacketization levels while supporting a large number of users. Our scheme outperforms MaN scheme in terms of rate per user, gains and subpacketization simultaneously. It can perform better than SPE scheme in terms of users supported, rate per user and subpacketization levels which are important design parameters for any coded caching scheme. Our scheme also supports a wide range of choices of KM/N as opposed to SPE scheme. We have shown that the schemes presented in the paper using resolvable designs from affine planes perform better than the MaN scheme for large memory sizes using the metric per-user-rate. This is the only class of resolvable designs that we could identify which is cross resolvable. It will be interesting to construct or identify new cross resolvable designs and study the performance of the resulting multi-access coded caching schemes.

\section*{Acknowledgment}
	This work was supported partly by the Science and Engineering Research Board (SERB) of Department of Science and Technology (DST), Government of India, through J.C. Bose National Fellowship to B. Sundar Rajan.

\appendix

\begin{center}
\bf{Proof of Correctness of the delivery algorithm}
\end{center}
The proof of correctness of the delivery phase given by {\bf Algorithm 1} is provided by the sequence of the following three lemmas.
\begin{lem} 
\label{lem5}
Let ${Y_{m}},\;m \in \mathcal{K}$ be the set of indices of subfiles accessible to the $m^{th}$ user. It is easily seen that $|Y_m|=\frac{vM'}{N}.$  The set of subfiles which is available with every user in $\mathcal{X}$, other than $m$ is $ \underset{t \in \mathcal{X}\setminus m} \cap  {Y_{t}},\;\forall m \in \mathcal{X}.$ Consider the transmission corresponding to the set $\mathcal{X}$ and a user $m\in \mathcal{X}$ as in {\bf Algorithm 1}. Then the following equality holds.
\begin{equation}
\label{fmequality}
f_{m} = \underset{t \in \mathcal{X}\setminus m} \cap  {Y_{t}},\;\forall m \in \mathcal{X}
\end{equation}
\end{lem}
\begin{IEEEproof} In \textbf{Algorithm 1} consider the combination of $2z$ blocks (caches)      $C_{1,i_1},C_{1,j_1},C_{2,i_2},C_{2,j_2},\dots,C_{z,i_z},C_{z,j_z}$, where $i_s, j_s \in [b_r] \text{ and } j_s \neq i_s, \forall s \in [z]$ and the $m^{th}$ user which has access to $z$ blocks $C_{1,a_1},C_{2,a_2},\dots,C_{z,a_z}$, where $a_s\in \{i_s,j_s\},\;s \in [z]$. The sequence of equations (2) to (10) in the next page constitute the proof for \eqref{fmequality}. 
\begin{figure*}
\label{figstar}
\begin{eqnarray}
&        \underset{t \in \mathcal{X}\setminus m} \cap  {Y_{t}} &= \bigcap_{\substack{
        l_s\;=\;\{i_s, j_s\},\;\forall s \;\in\; [z] \\ (l_1,l_2,\dots,l_z)\;\neq\;(a_1,a_2,\dots,a_z)}}\{C_{1,l_1}\cup C_{2,l_2}\cup\dots\cup C_{z,l_z}\}\\
&         &= \left\{\bigcap_{\substack{l_s\;=\;\{i_s, j_s\},\\\forall s\; \in\; [2,z]}}\{C_{1,e_1}\cup C_{2,l_2}\cup C_{3,l_3}\cup\dots\cup C_{z,l_z}\}\right\}\bigcap
        \left\{\bigcap_{\substack{l_s\;=\;\{i_s, j_s\},\\\forall s\;\in\;[2,z] \\ (l_2,l_3,\dots,l_z)\;\neq\\(a_2,a
        _3,\dots,a_z)}}\{C_{1,a_1}\cup C_{2,l_2}\cup C_{3,l_3}\cup\dots\cup C_{z,l_z}\}\right\} \nonumber \\   
&        &=\;C_{1,e_1}\bigcap\left\{\bigcap_{\substack{l_s\;=\;\{i_s, j_s\},\\ \forall s\; \in \;[2,z] \\ (l_2,l_3,\dots,l_z)\;\neq\\(a_2,a_3,\dots,a_z)}}\{C_{1,a_1}\cup C_{2,l_2}\cup C_{3,l_3}\cup\dots\cup C_{z,l_z}\}\right\}\\
&        & = \; C_{1,e_1}\bigcap 
        \left\{C_{1,a_1}\bigcup\left\{\bigcap_{\substack{l_s\;=\;\{i_s, j_s\},\\ \forall s\;\in\;[2,z] \\ (l_2,l_3,\dots,l_z)\;\neq\\(a_2,a_3,\dots,a_z)}}\{C_{2,l_2}\cup C_{3,l_3}\cup\dots\cup C_{z,l_z}\}\right\}\right\} \\ 
&        &= \;C_{1,e_1}\bigcap\left\{\bigcap_{\substack{l_s\;=\;\{i_s, j_s\},\\ \forall s\;\in\;[2,z] \\ (l_2,l_3,\dots,l_z)\;\neq\\(a_2,a_3,\dots,a_z)}}\{C_{2,l_2}\cup C_{3,l_3}\cup\dots\cup C_{z,l_z}\}\right\}\\
&        &= \;\{C_{1,e_1}\cap C_{2,e_2}\}\bigcap
        \left\{\bigcap_{\substack{l_s\;=\;\{i_s, j_s\},\\ \forall s\;\in;[3,z] \\ (l_3,l_4,\dots,l_z)\;\neq\\(a_3,a_4,\dots,i_z)}}\{C_{3,l_3}\cup C_{4,l_4}\cup\dots\cup C_{z,l_z}\}\right\} \\
&   & \mbox{~~(proceeding as to arrive at the previous step)} \nonumber \\
&        &= \;\{C_{1,e_1}\cap C_{2,e_2}\cap C_{3,e_3}\}\bigcap
        \left\{\bigcap_{\substack{l_s\;=\;\{i_s, j_s\},\\ \forall s\;\in\;[4,z] \\ (l_4,l_5,\dots,l_z)\;\neq\\(a_4,a_5,\dots,a_z)}}\{C_{4,l_4}\cup C_{5,l_5}\cup\dots\cup C_{z,l_z}\}\right\}\\
&        &\vdots \\
&        & = \;\{C_{1,e_1}\cap C_{2,e_2}\cap\dots\cap C_{z-1,e_{z-1}}\}\bigcap 
        \left\{\bigcap_{\substack{l_z\;=\;\{i_z, j_z\},\\(l_z)\;\neq\;(a_z)}}\{C_{z,l_z}\}\right\}\\
&        & = \;\{C_{1,e_1}\cap C_{2,e_2}\cap\dots\cap C_{z,e_{z}}\}=f_m. \\ \hline \nonumber
\end{eqnarray}

\end{figure*}
\end{IEEEproof}
\begin{lem}
\label{lem6}
 At the end of each transmission corresponding to set $\mathcal{X}$ in {\bf Algorithm 1}, each user is able to decode one sub-file.
\end{lem}
\begin{IEEEproof}
        Consider a user $m,\;m\in \mathcal{X}$ and any user $m',\;m'\neq m,\;m'\in\mathcal{X}$. The set $f_{m'}$ represents the set of subfiles which is available with every  user in $\mathcal{X}$, other than $m'$. From Lemma \ref{lem5} we know that the user $m$ has access to all the subfiles in $f_{m'}.$  Consider the transmission corresponding to set $\mathcal{X}$
        $$\underset {m \in \mathcal{X}} \oplus W_{d_{m},y_{m,s}},\; s \in [\mu_{z}]$$
        The $m^{th}$ user is able to get the subfile $ W_{d_{m},y_{m,s}}$
        from this transmission since  it has every other subfile $W_{d_{m'},y_{m',s}},\;\forall m'\in \mathcal{X},\;m'\neq m$.
\end{IEEEproof}
\begin{lem} 
\label{lem7}
At the end of all the transmissions in {\bf Algorithm 1} each user is able get all subfiles of the demanded file $W_{d_m},\;m\in \mathcal{K}.$ 
\end{lem}
\begin{IEEEproof}
In {\bf Algorithm 1} it can noted that there are in total $(b_r-1)^z$ transmissions from which the $m^{th}$ user gets $\mu_z(b_r-1)^z$ subfiles. Now consider a combination of $2z$ blocks (caches) given as $C_{1,i_1},C_{1,j_1},C_{2,i_2},C_{2,j_2},\dots,C_{z,i_z},C_{z,j_z}$, where $i_s, j_s \in [b_r] \text{ and } j_s \neq i_s, \forall s \in [z]$, and the $m^{th}$ user having access to $z$    blocks (caches), $C_{1,a_1},C_{2,a_2},\dots,C_{z,a_z}$, where $a_s\in\{i_s,j_s\},\;\forall s \in [z]$. We have
$$Y_m = C_{1,a_1}\cup C_{2,a_2}\cup\dots\cup C_{z,a_z}$$
and the $m^{th}$ user is able to receive subfile indices of the demanded file $W_{d_m}$ from the transmission corresponding to $2z$ caches (considered above) given by
        $$\bigcap_{\substack{l_s\;=\;\{i_s, j_s\},\;\forall s\;\in\;[z] \\ (l_1,l_2,\dots,l_z)\;\neq\;(a_1,a_2,\dots,a_z)}}\{C_{1,l_1}\cup C_{2,l_2}\cup\dots\cup C_{z,l_z}\} $$
which, using Lemma \ref{lem5}  is same as 
        $$\{C_{1,e_1}\cap C_{2,e_2}\cap\dots\cap C_{z,e_{z}}\}$$
where, $e_s = \{i_s,j_s\}\setminus a_s,\;\forall s \in [z]$. In order to find subfile indices that $m^{th}$ user get from all $(b_r-1)^z$ transmissions, we have to vary the value of $e_s$ such that $e_s\neq a_s,\;\forall s\in [z]$.\\
        The $m^{th}$ user is able to receive subfile indices given by
        $$\bigcup_{\substack{e_s\;=\;1,\\e_s\;\neq\;a_s\\\forall s\;\in\; [z]}}^{b_r}\{C_{1,e_1}\cap C_{2,e_2}\cap\dots\cap C_{z,e_{z}}\}.$$
In addition to this using $Y_m,$ the $m^{th}$ user gets the subfile indices shown in the sequence of expressions numbered (11) to (17). Notice that the last expression in (17) is the union of all the blocks in a parallel class. So from the property of resolvable designs  the above set is equal to the set containing all the subfile indices of all the files and therefore it also contains all the subfiles of the demanded file $W_{d_m}$.
\begin{figure*}
\begin{small}
\begin{eqnarray}
& & Y_m\bigcup\left\{\bigcup_{\substack{e_s\;=\;1,\\e_s\;\neq\;a_s\\\forall s\;\in\; [z]}}^{b_r}\{C_{1,e_1}\cap C_{2,e_2}\cap\dots\cap C_{z,e_{z}}\}\right\} \\  
& &                       =\left\{C_{1,a_1}\cup C_{2,a_2}\cup\dots\cup C_{z,a_{z}}\right\}\bigcup
                \bigcup_{\substack{e_s\;=\;1,\\e_s\;\neq\;a_s\\\forall s\;\in\; [z-1]}}^{b_r}\left\{C_{1,e_1}\cap C_{2,e_2}\cap\dots\cap \left\{\overset{b_r}{\underset{\substack{e_z=1\\e_z\;\neq\;a_z}}{\cup}}C_{z,e_{z}}\right\}\right\} \\ 
&  &      \supseteq\left\{C_{1,a_1}\cup C_{2,a_2}\cup\dots\cup C_{z-1,a_{z-1}}\right\}\bigcup\bigcup_{\substack{e_s\;=\;1,\\e_s\;\neq\;a_s\\\forall s\;\in\; [z-1]}}^{b_r}\left\{C_{1,e_1}\cap C_{2,e_2}\cap\dots\cap  C_{z,a_z}\right\}\bigcup
        \bigcup_{\substack{e_s\;=\;1,\\e_s\;\neq\;a_s\\\forall s\;\in\; [z-1]}}^{b_r}\left\{C_{1,e_1}\cap C_{2,e_2}\cap\dots\cap \left\{\overset{b_r}{\underset{\substack{e_z=1\\e_z\;\neq\;a_z}}{\cup}}C_{z,e_{z}}\right\}\right\} \nonumber \\ 
&  &      =\left\{C_{1,a_1}\cup C_{2,a_2}\cup\dots\cup C_{z-1,a_{z-1}}\right\}\bigcup
        \bigcup_{\substack{e_s\;=\;1,\\e_s\;\neq\;a_s\\\forall s\;\in\; [z-1]}}^{b_r}\left\{C_{1,e_1}\cap C_{2,e_2}\cap\dots\cap\left\{\overset{b_r}{\underset{e_z=1}{\cup}}C_{z,e_{z}}\right\}\right\}\\
&   &     =\left\{C_{1,a_1}\cup C_{2,a_2}\cup\dots\cup C_{z-1,a_{z-1}}\right\}\bigcup
        \bigcup_{\substack{e_s\;=\;1,\\e_s\;\neq\;a_s\\\forall s\;\in\; [z-1]}}^{b_r}\left\{C_{1,e_1}\cap C_{2,e_2}\cap\dots\cap C_{z-1,e_{z-1}}\right\} \\ \nonumber
&    &    \mbox{(Continuing with similar steps  we get),}\\
&     &   \supseteq\left\{C_{1,a_1}\cup C_{2,a_2}\cup\dots\cup C_{z-2,a_{z-2}}\right\}\bigcup
        \bigcup_{\substack{e_s\;=\;1,\\e_s\;\neq\;a_s\\\forall s\;\in\; [z-2]}}^{b_r}\left\{C_{1,e_1}\cap C_{2,e_2}\cap\dots\cap C_{z-2,e_{z-2}}\right\} \\ \nonumber
&    &  \vdots \\ 
&    &    \supseteq\left\{C_{1,a_1}\cup C_{2,a_2}\right\}\bigcup
        \left\{\bigcup_{\substack{e_s\;=\;1,\\e_s\;\neq\;a_s\\\forall s\;\in\; [2]}}^{b_r}\left\{C_{1,e_1}\cap C_{2,e_2}\right\}\right\} \\
&    &    \supseteq\left\{C_{1,a_1}\right\}\bigcup
        \left\{\bigcup_{\substack{e_1\;=\;1,\\e_1\;\neq\;a_1}}^{b_r}C_{1,e_1}\right\} \\ \hline \nonumber 
\end{eqnarray}
\end{small}
\end{figure*}
\end{IEEEproof}


\end{document}